\documentclass[default]{aastex631}

\newcommand{\fo}{\ensuremath{f^\parallel}}
\newcommand{\fe}{\ensuremath{f^\perp}}
\newcommand{\pr}{\ensuremath{P_{\rm r}}}
\newcommand{\px}{\ensuremath{P_X}}
\newcommand{\pq}{\ensuremath{P_Q}}
\newcommand{\pu}{\ensuremath{P_U}}

\newcommand{\ainv}{\ensuremath{\alpha_{0}} }
\newcommand{\amin}{\ensuremath{\alpha_{\rm{min}}} }
\newcommand{\amax}{\ensuremath{\alpha_{\rm{max}}} }

\newcommand{\pmax}{\ensuremath{P_{\rm{max}}} }
\newcommand{\pmin}{\ensuremath{P_{\rm{min}}} }

\begin{document}

\title{Extreme Negative Polarisation of New Interstellar Comet 3I/ATLAS}

\author[0000-0002-6610-1897]{Zuri Gray}
\affiliation{Department of Physics, PO Box 64, FI-00014 University of Helsinki, Finland}

\author[0000-0002-7156-8029]{Stefano Bagnulo}
\affiliation{Armagh Observatory \& Planetarium, College Hill, Armagh, BT61 9DG, UK}

\author[0000-0002-4516-459X]{Galin Borisov}
\affiliation{Institute of Astronomy and National Astronomical Observatory, Bulgarian Academy of Sciences, 72 Tsarigradsko Chaussée Blvd., BG-1784 Sofia, Bulgaria}
\affiliation{Armagh Observatory \& Planetarium, College Hill, Armagh, BT61 9DG, UK}

\author[0000-0002-8122-3606]{Yuna G. Kwon}
\affiliation{Caltech/IPAC, 1200 E California Blvd, MC 100-22, Pasadena, CA 91125, USA}

\author[0000-0002-6645-334X]{Alberto Cellino}
\affiliation{INAF -- Osservatorio Astrofisico di Torino, I-10025 Pino Torinese, Italy}

\author[0000-0002-9321-3202]{Ludmilla Kolokolova}
\affiliation{Department of Astronomy, University of Maryland, College Park, MD 20742-2421, US}

\author[0000-0002-8910-1021]{Rosemary C. Dorsey}
\affiliation{Department of Physics, PO Box 64, FI-00014 University of Helsinki, Finland}

\author[0000-0002-8418-4809]{Grigori Fedorets}
\affiliation{Finnish Centre for Astronomy with ESO, University of Turku, FI-20014 Turku, Finland}
\affiliation{Department of Physics, PO Box 64, FI-00014 University of Helsinki, Finland}

\author[0000-0002-5624-1888]{Mikael Granvik}
\affiliation{Department of Physics, PO Box 64, FI-00014 University of Helsinki, Finland}
\affiliation{Asteroid Engineering Laboratory, Lule\r{a} University of Technology, Box 848, SE-98128 Kiruna, Sweden}

\author[0000-0002-9870-123X]{Eric MacLennan}
\affiliation{Department of Physics, PO Box 64, FI-00014 University of Helsinki, Finland}

\author[0000-0002-5138-3932]{Olga Mu\~{n}oz}
\affiliation{Instituto de Astrof\'{\i}sica de Andaluc\'{\i}a, CSIC, Glorieta de la Astronomia s/n, E-18008 Granada, Spain\\}

\author[0000-0002-4278-1437]{Philippe Bendjoya}
\affiliation{Université Côte d’Azur, Laboratoire Lagrange, OCA, CNRS UMR 7293, Nice, France}

\author[0000-0002-6509-6360]{Maxime Devogèle}
\affiliation{ESA NEO Coordination Centre, Largo Galileo Galilei 1, I-00044 Frascati (RM), Italy}

\author[0000-0001-8694-9038]{Simone Ieva}
\affiliation{INAF -- Osservatorio Astronomico di Roma, 00078 Monte Porzio Catone (RM), Italy}

\author[0000-0001-7403-1721]{Antti Penttil\"a}
\affiliation{Department of Physics, PO Box 64, FI-00014 University of Helsinki, Finland}

\author[0000-0001-8058-2642]{Karri Muinonen}
\affiliation{Department of Physics, PO Box 64, FI-00014 University of Helsinki, Finland}

% \author{Hermann Boehnhardt} % ORCID and affiliation?
% \affiliation{Armagh Observatory \& Planetarium, College Hill, Armagh, BT61 9DG, UK}

% \author[]{}
% \affiliation{}

\begin{abstract}
We present the first polarimetric observations of the third discovered interstellar object, 3I/ATLAS (C/2025 N1), obtained pre-perihelion with FORS2/VLT, ALFOSC/NOT, and FoReRo2/RCC, over a phase angle range of $7.7-22.4^\circ$. This marks the second ever polarimetric study of an interstellar object, the first distinguishing 2I/Borisov from most Solar System comets by its higher positive polarisation. Our polarimetric measurements as a function of phase angle reveal that 3I is characterised by an deep and narrow negative polarisation branch, reaching a minimum value of $-2.7\%$ at phase angle $7^\circ$, and an inversion angle of $17^\circ$---a combination unprecedented among asteroids and comets, including 2I/Borisov. At very small phase angles, the extrapolated slope of the polarisation phase curve is consistent with that of certain small trans-Neptunian objects and Centaur Pholus, consistent with independent spectroscopic evidence for a red, possibly water-ice-bearing object. Imaging confirms a diffuse coma present from our earliest observations, though no strong polarimetric features are spatial resolved. These findings may demonstrate that 3I represents a distinct type of comet, expanding the diversity of known interstellar bodies. 

\end{abstract}

\keywords{Polarimetry (1278), Interstellar Objects (52)}
% ===================================================================================================
% ============================================ SECTION 1 ============================================
\section{Introduction}
\label{Intro}
% (Note: remember to update references from arXiv to journals when they are published)

% ...General introduction to 3I...
The object 3I/ATLAS (C/2025 N1, initially designated A11pl3Z; hereafter referred to as 3I) was discovered by the ATLAS survey on July 1st 2025 and was quickly recognised to follow a retrograde, hyperbolic orbit with an eccentricity of $\sim6$ and an inclination exceeding $175^\circ$, securely establishing its extrasolar origin \citep{seligman2025arXiv}. It is only the third interstellar object (ISO) discovered to date, after 1I/'Oumuamua (hereafter 1I) and 2I/Borisov (hereafter 2I). The first two ISOs displayed radically different physical properties: 1I (discovered in 2017), an unexpectedly red, elongated, tumbling object, displayed non-gravitational accelerations without detectable activity despite deep searches \citep[e.g.,][]{meech2017Natur,micheli2018Natur}, while 2I (discovered in 2019) was an active comet with gas species and dust properties broadly consistent with Solar System comets, though with an unusually high CO abundance \citep[e.g.,][]{guzik2020NatAs, opitom2021AA} and atypical polarimetric properties \citep{bagnulo2021Natur}. Thus, with 1I appearing asteroid-like but enigmatic, and 2I mostly resembling a familiar comet, the arrival of 3I offers the possibility of sampling yet another member of the diverse ISO population, and of testing whether its dust and activity reflect known objects or not. Due to their hyperbolic orbits, ISOs only appear for a single passage through the Solar System; capture into a bound orbit \citep{hands2020,dehnen2022} or re-encounter with the Sun where a second orbit may be observed are extremely unlikely possibilities. Hence, the time between an ISO's discovery and its last observable moment post-perihelion is the critical period for characterising an ISO.

% ..... what do we know about 3I so far according to current publications?
Multi-technique observations---including astrometry, photometry, and spectroscopy---are being used to refine 3I's orbit, detect volatile outgassing, and characterise the dust and gas components of the coma for direct comparison with other Solar System objects. Early results show that 3I is quite red, similar to D-type asteroids, most likely due to an active dust coma in the absence of volatile gas sublimation near its discovery \citep{seligman2025arXiv, opitom2025arxiv, bolin2025mnras, belyakov2025rnaas, kareta2025arxiv, chandler2025arxiv}. Unsurprisingly, 3I hence displays little lightcurve variability \citep{seligman2025arXiv, kareta2025arxiv, chandler2025arxiv,delaFuenteMarcos2025AA}. Kinematic analysis of its inbound velocity compared to the expected velocity distribution of the galactic population suggest an origin distinct from either 1I or 2I, likely originating in an older, lower-metallicity star system \citep{hopkins2025arxiv, taylor2025arxiv, delaFuenteMarcos2025AA}. Specifically, \citet{hopkins2025arxiv} predicted that 3I would likely become more active closer to the water-ice line due to an estimated large fraction of mass as water; a prediction supported by recent indirect and direct detections of water-ice emission \citep{yang2025arxiv, xing2025arxiv} at atypically large heliocentric distances $r_h\geq3.51$~au. Infrared observations by the JWST at $r_h =3.3$~au reveal a CO$_2$ dominated coma, with a CO$_2$/H$_2$O among the highest ever recording at such a heliocentric distance, as well as the presence of H$_2$O, CO, OCS, water-ice and dust \citep{cordiner2025}. While already displaying similar cometary characteristics to 2I, 3I is the largest ISO ever discovered, with HST observations constraining its radius to $0.16 \leq r \leq 2.8$~km \citep{jewitt2025arxiv}, assuming an albedo $p_V=0.04$. The detailed characterisation of 3I currently ongoing by the astronomical community is vital for informing our expectations of interstellar asteroids and comets from different origins within the galaxy, as well as placing 1I and 2I into context of the broader ISO population \citep{flekkoy2023MNRAS}. 

% .....seway into polarimetry...
While spectroscopy and imaging help constrain composition and morphology, polarimetry uniquely probes physical properties of the object---such as optical constants, grain size distribution, and dust shape and structure in the case of a cometary coma or surface regolith---that are not easily inferred from other techniques. The degree of linear polarisation $P_r$ of sunlight scattered by cometary dust particles is measured as the difference between the fluxes perpendicular and parallel to the scattering plane (Sun-comet-observer plane), normalised by their sum, and is highly sensitive to the solar phase angle $\alpha$ (Sun-comet-observer angle). Because of this definition, polarisation is positive when the polarisation direction is oriented perpendicular to the scattering plane, and negative when parallel. Overall, Solar System objects show similar phase angle dependence \pr$(\alpha)$: negative polarisation at small phase angles ($\lesssim 20^\circ$), and positive at larger phase angles, increasing nearly linearly until reaching a maximum at $\simeq 90-100^\circ$. The detailed shape of this curve varies according to each object: the minimum and maximum values of polarisation (\pmax and $\pmin$), their corresponding phase angles (\amax and $\amin$), the so-called inversion angle \ainv where the polarisation switches sign, and the slope of the curve at this angle $h=\Delta P/\Delta \alpha$. Combined with photometric and spectroscopic data, these six parameters provide a framework for theoretical studies to constrain the physical properties of the scattering media. For example, using large porous particles with highly absorbing inclusions in a laboratory setting, \citet{munoz2020} was able to reproduce both the OSIRIS/Rosetta phase function and polarimetric phase curve obtained via ground-based observations of comet 67P/Churyumov-Gerasimenko. 

%.....outline polarimetry of "regular" comets VS 2I 
In the case of cometary atmospheres, dust properties (hence, polarisation) can vary throughout its apparition according to heliocentric distance $r$ and cometocentric distance $\rho$ (and therefore, aperture radius). Despite accounting for these variables, consistent trends still emerge. Studies show that most comets generally fall into one of two broad polarimetric classes defined by their \pmax values: high-polarisation comets, with $\pmax \simeq 25-30\%$ at $\amax \simeq95^\circ$; and low-polarisation comets, with $\pmax \simeq 8-22\%$ at $\amax \simeq 90^\circ$ \citep{kiselev2015}. A handful of exceptional objects deviate significantly from these classes. The best-known case is C/1995 O1 (Hale-Bopp), which shows polarisation values about $4\%$ higher than typical high-polarisation comets at phase angles $\gtrsim35^\circ$, leading to the definition of a distinct "Hale-Bopp-like" polarimetric class \citep{hadamcik2003}. For many years, Hale-Bopp remained unique in this regard, until the appearance of 2I. VLT polarimetric observations of 2I by \citet{bagnulo2021Natur} revealed a phase dependence remarkably similar to that of Hale-Bopp, despite spectroscopic and photometric studies finding it otherwise consistent with ordinary Solar System comets. These data thus established 2I as the first ISO with polarimetric measurements, and placed it in the exceptional Hale-Bopp class. 

The arrival of 3I provides a new opportunity to extend such studies. In this letter, we present the first polarimetric observations of 3I which probe the poorly explored negative polarisation branch of an ISO.

% ===================================================================================================
% ============================================ SECTION 2 ============================================
\section{Observations and Analysis}\label{Sect_Observations}
We obtained 18 pre-perihelion imaging polarimetric observations of 3I using three different instruments between July 17 and August 28, 2025, corresponding to a phase angle range of $7.7$--$22.4^{\circ}$ and heliocentric distance range of $3.9$--$2.6$~au. Observations were conducted in the R-band using FORS2 at the ESO VLT (8.2\,m), ALFOSC at the NOT (2.56\,m), and FoReRo2 at the BNAO Rozhen (2\,m), while three additional V-band observations were acquired with FORS2. A summary of the observing log is presented in Table~\ref{tableLog}. We obtained one earlier observation with ALFOSC on July 11, 2025, (at phase angle $5.7^\circ$) but were unable to obtain a reliable measurement due to the dense star field at the time of the observation. 

For the analysis of our data, we performed aperture polarimetry using a $\sim2000$ km radius circular aperture centred on the photocentre. We opted for this small aperture size to avoid contamination from background field stars, of which there were many in the earliest epochs. This aperture method is useful for comparing the bulk properties of 3I to that of other objects, but loses spatial information due to the averaging of integrated signal within the aperture. We overcame this issue by constructing polarimetric maps and deep imaging maps from our FORS2 data. Full details of the data reduction procedures for these instruments are provided, for example, in \citet{gray2024PSJ}, \citet{gray2024MNRAS}, and \citet{borisov2015PSS}. We outline the basic principles in the Appendix. 

% ============================================ SECTION 3 ============================================

\section{Results}\label{Sect_Results}

\subsection{Aperture Polarimetry}
\begin{figure*}
\begin{center}
\includegraphics[width=0.75\textwidth]{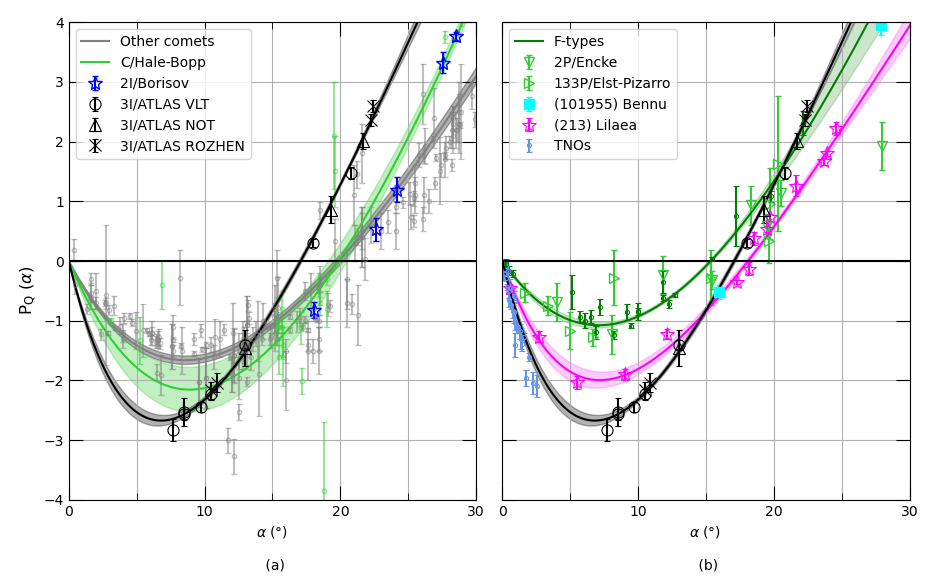}
\end{center}
\caption{\label{Figure1} Polarisation versus phase angle of 3I compared to various other objects. Solid lines with shaded areas represent best fit curves calculated according to Eq.~\ref{Eq_Fit} with $\pm 1\sigma$ uncertainty. Left: "other comets" refers to typical Solar System comets, such as 9P/Tempel 1, 22P/Kopff, 47P/Ashbrook-Jackson, 67P/Churyumov-Gerasimenko, etc., sourced from the Database of Comet Polarimetry \citep{kiselev2017}. Right: F-type curve is calculated using (302) Clarissa, (419) Aurelia, and (704) Interamnia data, sourced from the Asteroid Polarimetric Database \citep{lupishko2019}; TNOs datapoints include Huya, Ixion, 1999 DE9, and Varuna, as well as Centaur Pholus, sourced from the TNO Polarimetric Database \citep{belskaya2013}. The 3I data in the right image is as labelled in the left. All other datapoints are as labelled in the legends. All data are in R-band, except for (213) Lilaea, for which only V-band data exists.} 
\end{figure*}

The values of the linear polarisation measurements at each epoch are listed in Table~\ref{tableLog} and plotted as a function of phase angle in Fig.~\ref{Figure1}, together with data for various Solar System bodies. Our observations show that 3I exhibits expected polarimetric behaviour, showing negative polarisation at small phase angles and a linear increase of polarisation with increasing phase angle. In order to compare 3I to other objects, we fit the linear-exponential empirical model suggested by \citet{muinonen2009} to the data:
% ------------------------------------------- EQUATION 3 -------------------------------------------
\begin{equation}
  \pr(\alpha) = A \left({\rm e}^{-(\alpha/B)} - 1\right) + C\,\alpha
\label{Eq_Fit}
\end{equation}
% --------------------------------------------------------------------------------------------------
where $\alpha$ is the phase angle, and $A$, $B$, and $C$ are free parameters which shape the curve. The best-fit parameters of 3I are given in Table~\ref{tableCurve}.

According to our R-band data, 3I is characterised by an inversion angle of $\ainv = 17.1 \pm 0.1^\circ$---a particularly small value compared to the typical value of $20$--$22^\circ$ of Solar System comets, as well as that of the Hale-Bopp-like group to which 2I (characterised by $\ainv\simeq20.5^\circ)$ belongs. With values $\pmin = -2.7\%$ at $\amin = 7^\circ$, the negative polarisation branch of 3I is notably deeper and shifted to smaller phase angles than that observed for these comet classes, as shown in Fig.~\ref{Figure1}(a). The slope at the inversion angle is another key diagnostic parameter, with empirical studies showing that asteroids exhibit a direct correlation between slope and geometric albedo---the steeper the slope, the lower the albedo---known as Umov's law \citep{zellner1976,cellino2015}. From our 3I curve, we measure a polarimetric slope of $h=0.42 \%$~deg$^{-1}$, a value somewhat extreme among the typical $0.2$--$0.4\%$~deg$^{-1}$ range for Solar System comets, but comparable to the $0.45\%$~deg$^{-1}$ slope of 2I. If 3I were an asteroid or we were observing the comet nucleus directly, such a slope would correspond to an extremely low albedo ($\rho_v \simeq 0.04$). However, the empirical albedo-polarisation slope relationship used for asteroids cannot be strictly applied to cometary comae due to the difference in light scattering physics between densely packed particulate surfaces versus diffuse particulate clouds \citep{zubko2011}. Either way, the combination of a shallow inversion angle and deep negative polarisation branch is unprecedented among comets observed to date.

\begin{table}
% \centering
% \begin{center}
\caption{\label{tableLog} Results of aperture polarimetric measurements of 3I using aperture radius of approximately 2000 km. In cases where a column/row parameter is not listed, its value is equal to that in the previous row. This is except for \pu, where a missing value implies the instrument position angle is already rotated to the scattering plane--see Appendix.} 
\begin{center}
% \centering
\begin{tabular}{cccccr@{$\pm$}lr@{$\pm$}l}
\hline\hline
Date & Instrument & $\alpha$  & $r$ & Filter & \multicolumn{2}{c}{\pq} &   \multicolumn{2}{c}{\pu} \\
     &        & (deg)     & (au)&      & \multicolumn{2}{c}{$(\%)$} & \multicolumn{2}{c}{$(\%)$} \\
\hline \hline
2025-Jun-18 & FORS2 &  8.5 & 3.92 & R & -2.53 & 0.23 &  0.18 & 0.26 \\
2025-Jun-19 &  --   &  --  &   -- & V & -2.87 & 0.14 & \multicolumn{2}{c}{---} \\
    --      &  --   &  --  &   -- & R & -2.57 & 0.09 & \multicolumn{2}{c}{---} \\
2025-Jul-17 &  --   &  7.7 & 3.98 & R & -2.83 & 0.19 &  0.12 & 0.24 \\
     --     &  --   &  --  &   -- & V & -2.72 & 0.16 & \multicolumn{2}{c}{---} \\
2025-Jul-22 &  --   &  9.7 & 3.82 & R & -2.44 & 0.08 & \multicolumn{2}{c}{---} \\
2025-Jul-23 &FoReRo2& 10.5 &   -- & R & -2.17 & 0.14 &   \multicolumn{2}{c}{---} \\
2025-Jul-24 & FORS2 & --   & 3.75 & R & -2.23 & 0.09 & \multicolumn{2}{c}{---} \\
    --      &FoReRo2& 10.8 &   -- & R & -2.03 & 0.16 &   \multicolumn{2}{c}{---} \\
2025-Jul-29 & ALFOSC& 13.0 & 3.56 & R & -1.46 & 0.30 &  0.01 & 0.30 \\
2025-Jul-30 & FORS2 &  --  &   -- & R & -1.40 & 0.07 & \multicolumn{2}{c}{---} \\
2025-Aug-12 &  --   & 18.0 & 3.14 & R &  0.30 & 0.07 & \multicolumn{2}{c}{---} \\
2025-Aug-15 & ALFOSC& 19.3 & 3.01 & R &  0.86 & 0.23 & -0.26 & 0.23 \\
2025-Aug-21 & FORS2 & 20.8 & 2.85 & R &  1.48 & 0.10 & -0.18 & 0.10 \\
       --   & --    &  --  & --   & V &  1.15 & 0.11 & \multicolumn{2}{c}{---} \\
2025-Aug-24 & ALFOSC& 21.7 & 2.73 & R & 2.18 & 0.14 & 0.01 & 0.14 \\ 
2024-Aug-27 &FoReRo2& 22.3 & 2.64 & R & 2.37 & 0.10 & 0.03 & 0.10 \\
2024-Aug-28 &  --   & 22.4 & 2.61 & R & 2.59 & 0.10 & 0.23 & 0.10 \\
\hline
\end{tabular}
\end{center}
\end{table}

% -------------------------------------------- FIGURE 1 --------------------------------------------
% -------------------------------------------------------------------------------------------------
\begin{table}
    %\centering
    \caption{R-band polarimetric phase curve parameters of 3I according to best fit.}
    \centering
    \begin{tabular}{lr@{$\pm$}lr@{$\pm$}lr@{$\pm$}lr@{$\pm$}l}
        \hline \hline
        Object & \multicolumn{2}{c}{\pmin ($\%$)} & \multicolumn{2}{c}{\amin ($^{\circ}$)} & \multicolumn{2}{c}{\ainv ($^{\circ}$)} & \multicolumn{2}{c}{$h$ ($\%$~deg$^{-1}$)} \\
        \hline
        3I/ATLAS & -2.67 & 0.08 & 6.79 & 1.34 & 17.08 & 0.12 & 0.416 & 0.008 \\
        \hline 
    \end{tabular}
    \label{tableCurve}
\end{table}

\subsection{Imaging polarimetry}
\begin{figure*}
\begin{center}
\includegraphics[width=\textwidth]{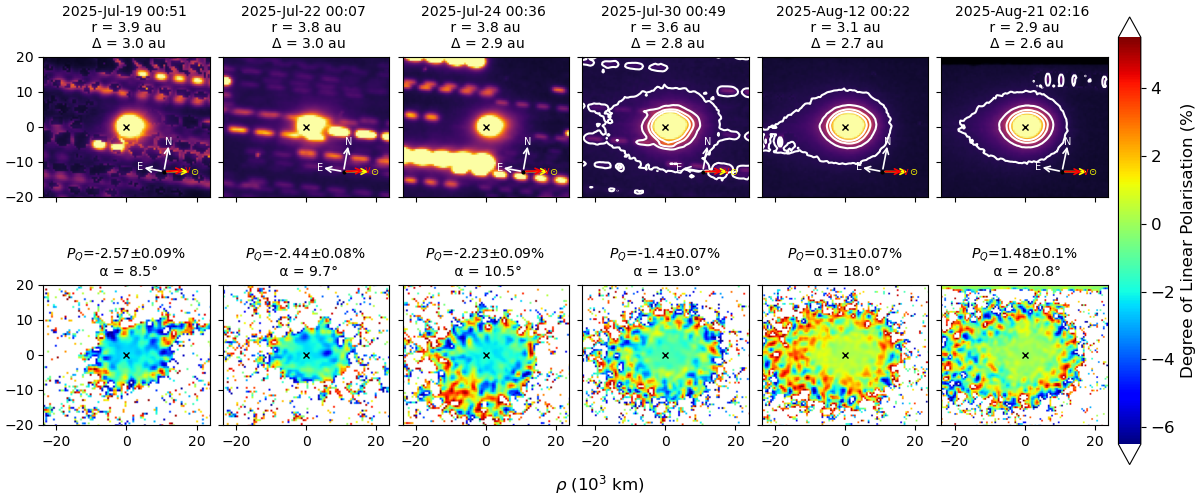}
\end{center}
\caption{\label{Figure2} Deep imaging (top) and polarimetric (bottom) maps of 3I from a subset of VLT observations. The colour scale in the imaging maps does not reflect the absolute brightness of the comet. We display isophotes in the last three epochs only due to the large number of background stars in the first images. The colour of each pixel in the polarimetric maps represents the value of polarisation, as shown in the scale to the right. Pixels with values outside this colour scale are given in white and are considered noise. The arrows indicate the directions towards the celestial north and east, as well as the Sun-comet direction and velocity direction of the comet (which mostly overlap) projected onto the sky. X marks the comet photocentre.} 
\end{figure*}

Fig.~\ref{Figure2} presents R-band deep imaging and polarimetric maps for a subset of our FORS2 data. For the last three epochs, we overlap isophotes to highlight the morphology of the extended coma, but we omit them in the first three images due to the high density of background stars. In these early frames, the photocentre of 3I passes close to several field stars, illustrating the rationale for adopting a small aperture for the aperture polarimetric measurements. We verified that this aperture was sufficiently small to avoid contamination from the background stars---i.e. while the polarimetric map of the extended coma in the July 22 epoch (in the projected direction south of the photocentre) is affected by a nearby star, we remain confident that our aperture polarimetric measurement remains intact. We include these maps despite the dense star field for completeness.

A weak diffuse coma is apparent from our earliest datasets, which grew steadily as the heliocentric distance decreased with time. No clear tail was visible at these early epochs---likely projected behind the comet, hidden from our point of view due to the observing geometry---but later emerged as an extension of the coma in the anti-solar direction. The absence of a prominent tail in early observing epochs is consistent with reports, for example, by \citet{kareta2025arxiv} and \citet{santana2025arXiv}. The polarimetric maps reveal no significant spatial structures or variations across the coma and tail. This is in line with recent polarimetric studies of other comets, including 2I \citep{bagnulo2021Natur}, Oort-cloud comet C/2017 K2 (PANSTARRS) \citep{kwon2024}, and Jupiter Family Comet 67P/Churyumov-Gerasimenko \citep{gray2024MNRAS}. Continued monitoring of 3I with imaging and polarimetric mapping in future epochs will be crucial to track the evolution of its coma and better understand its dust properties and activity in comparison to other comets.

% -------------------------------------------- FIGURE 2 --------------------------------------------
% -------------------------------------------------------------------------------------------------

\section{Discussion}
The inversion and minimum angles of 3I approach those of classic F-type asteroids ($\ainv \simeq 15^\circ$, $\amin \simeq 7^\circ $) as defined by \citet{tholen1984}. Though limited, polarimetric observations of cometary nuclei, namely 2P/Encke \citep{jewitt2004,boehnhardt2008} and 133P/Elst-Pizarro \citep{bagnulo2010}, likewise show low inversion angles similar to that of F-type asteroids. Moreover, (4015) Wilson-Harrington, an object originally classified as an F-type asteroid upon its discovery \citep{bowell1992}, was later found to exhibit cometary activity. A lack of further activity detections in subsequent observations, however, suggest that Wilson-Harrington may be a comet becoming extinct. Intriguingly, more recent work has strengthened this connection: (101955) Bennu, the target of OSIRIS-REx sample return mission, was found to display F-type-like polarimetric properties \citep{cellino2018} despite previously being classified as a B-type according to the \citet{busbinzel2002} taxonomic classification system. Analysis of the first data collected in-situ by the spacecraft later revealed that Bennu is a low-albedo asteroid with a surface rich in hydrated minerals and with moderate surface activity, and even with evidence of past water flow within the asteroid \citep{hergenrother2020,ishimaru2024}, reinforcing the idea of possible cometary nature of members of the F-class. These pieces of evidences suggest an inherent link between F-type asteroids and cometary nuclei and, thus, it has been proposed that a low inversion angle could be diagnostic of a cometary surface.

Some known asteroids illustrate that a low inversion angle does not always coincide with the typical depth of the negative branch seen in F-type asteroids. For example, (213) Lilaea, originally classified as F-type based on albedo and spectrophotometric data, is characterised by polarisation phase curve parameters $\ainv \simeq 18.1^\circ$, $\amin \simeq 7.1^\circ$ and $\pmin \simeq -2\%$---a negative polarisation branch roughly twice as deep as that of typical F-types, as shown in Fig.~\ref{Figure1}(b). In contrast, (269) Justitia (not listed in the figure to avoid overcrowding) exhibits $\ainv \simeq 15^\circ$, $\amin \simeq 6.5^\circ$, and $\pmin\simeq-0.6\%$---about half the typical depth \citep{gilhutton2014}. Spectral studies of (269) Justitia reveal an unusually steep red slope, redder than that of D-type asteroids, and closer to that of some TNOs \citep{hasegawa2021,hasegawa2022,humes2024A}. The peculiarity of this particular object has also been noted by \citet{cellino2020} and \citet{mahlke2022}, with the latter assigning (269) Justitia to the rare and relatively new Z-type taxonomic class. Overall, these examples suggest that objects with inversion angles around $15^\circ$ often prove to be anomalous and intrinsically interesting cases, worthy of further study. 

Despite sharing a small inversion angle with these unique cases, the depth of 3I's negative polarisation branch still sets it apart entirely. We report the first clear example of a planetary body whose polarimetric behaviour is dominated by such deep and narrow negative polarisation branch. Such behaviour has never been observed among Solar System objects: not in comets, which typically show shallower negative branches and larger inversion angles, nor in asteroids, even those belonging to rare and unusual taxonomic classes. 

At very small phase angles ($\alpha \leq 2^\circ$), the extrapolated fit of 3I most closely resembles the steep polarisation curves observed for small trans-Neptunian objects (TNOs; diameters 310--670 km), Huya, Ixion, Varuna, and 1999 DE$_9$, as well as Centaur Pholus \citep{bagnulo2006A&A,belskaya2010} (see Fig.~\ref{Figure1}(b)). These objects, only observable at small phase angles from ground-based observatories, are characterised by an extremely rapid increase in polarisation (absolute value), showing a much steeper slope at 0--2$^\circ$ compared to other objects. It's worth noting, however, that this polarimetric behaviour is not universal among all TNOs and Centaurs---most other objects show much slower variations in polarisation with phase angle. The common trait among these four TNOs is a deeper negative branch compared to larger and methane-rich TNOs, as well as belonging to the same IR taxonomic class \citep{fulchignoni2008} showing moderately red, featureless visible to near-IR spectra with weak to strong detections of water ice \citep{barucci2011,alvarez2007}. 

The similarity between 3I's steep polarimetric slope at small phase angles and that of certain TNOs is in line with independent spectroscopic and colour measurements which show that 3I appears redder than Solar System comet and D-type asteroids \citep{opitom2025arxiv,bolin2025mnras,delaFuenteMarcos2025AA,alvarez2025AA,santana2025arXiv}, possibly closer to the surfaces of some TNOs and Centaurs. Further, the predicted presence of water ice \citep{hopkins2025arxiv} was detected both indirectly and directly via spectroscopic measurements \citep{yang2025arxiv,xing2025arxiv}. 

Among theoretical studies, numerical simulations suggest that the observed polarimetric behaviour of TNOs and Centaurs arises from a surface mixture of bright and dark scatterers, and that the different polarimetric behaviours (i.e. fast and slow variation with phase angle) among TNOs and Centaurs can be explained via different ratios of these scatterers \citep[e.g.,][]{bagnulo2006A&A,belskaya2008}. Laboratory studies show that even small amounts of bright grains added to dark powders significantly enhances negative polarisation \citep{levasseurregourd2015}. In particular, laboratory experiments by \citet{dougherty1994} and \citet{poch2018} demonstrate that thin frost layers of water ice on dark surfaces produce deep minima shifted towards small phase angles, together with a reduced inversion angle---closely resembling the behaviour observed for some TNOs and 3I. Of course, in the case of 3I, our polarisation measurements are not probing the consolidated regolith surface but rather a diffuse coma of dust and ice grains. Interestingly, \citet{dougherty1994} found a similar minimum at small phase angles, but with reduced depth, together with a small inversion angle when repeating the experiments for a diffuse cloud of water-ice particles over a dark surface. Light scattering models of large cometary particles by \citet{ivanova2023} show that reproducing a deep polarisation minimum requires the inclusion of ice particles in highly porous particles. Their model, which treats grains as aggregates of sub-micron monomers with silicate-organic cores (50/50 mixture) and water- and CO$_2$-ice mantles (90/10 mixture), demonstrates that both particle size and porosity play key roles in shaping the polarimetric phase curves: larger particles shift \amin toward smaller phase angles (see also \citealt{frattin2022}), while \pmin~increases with porosity but exhibits non-linear dependence on particle size. With these studies in mind, the extreme depth of the negative polarisation branch of 3I hints to a mixture of icy and dark material, though the exact properties of the particles are still uncertain. The picture may become clearer with further observations as well as theoretical studies combining polarimetry with results from other observational techniques.

We caution that the comparison between 3I and TNOs remain tentative. Our measurements of \pmin and \amin correspond to the upper limit only since our data does not show the expected turn-over towards zero polarisation for small phase angles. Further, the lack of data at $\alpha < 7^\circ$ for 3I and $\alpha>2^\circ$ for TNOs makes a full comparison difficult. Observations of 3I after perihelion, when it will be visible in the phase angle range $\sim0-30^\circ$ will be critical to confirm its polarimetric behaviour at small phase angles and explore the genuinity of the parallels with that of TNOs and Centaurs.

\section{Summary and Conclusions}
In this letter, we present pre-perihelion imaging polarimetric observations of the third discovered interstellar comet, 3I/ATLAS. Our measurements establish 3I as uniquely distinct, compared to both solar system bodies and 2I, the only other interstellar object observed polarimetrically. 

3I is characterised by an extremely deep and narrow negative polarisation branch, with parameters $\pmin = -2.77 \pm 0.11\%$, $\amin= 6.41 \pm 1.27^\circ$, $\ainv = 17.05 \pm 0.12^\circ$, and $h = 0.40 \pm 0.02 \%$~deg$^{-1}$. This polarimetric behaviour is significantly different from all known comets (either interstellar or bound to our Solar System), not fitting into either the high- or low-polarisation comet categories, nor that of the unique category occupied by just Hale-Bopp and 2I. While 3I's inversion angle resembles that of rare F-type asteroids and cometary nuclei, the depth of its negative polarisation branch is almost twice as large. Extrapolation to very small phase angles indicate a steep polarisation slope between phase angles 0--2$^\circ$, reminiscent of certain small TNOs and Centaurs. While this finding is tentative due to a lack of data at small phase angles, it is consistent with the red spectral slope and evidence of water-ice detections in various independent studies. 

Imaging confirms the presence of a diffuse coma present from the earliest epochs which grows steadily with decreasing heliocentric distance. Our polarimetric maps reveal no significant spatial variations, suggesting homogeneous polarisation of the coma. Thus, our polarimetric measurements most likely probe the dust in the coma rather than the nucleus surface \citep[in agreement with][at a similar epoch]{jewitt2025arxiv}. Direct comparisons to atmosphereless bodies must be made with caution since the underlying light scattering physics differ fundamentally. Still, we explore 3I's similarity to all types of Solar System objects for completeness as our knowledge of the object continues to evolve. Based on theoretical studies of cometary particles \citep{ivanova2023}, the depth of the negative branch and small \amin~value for 3I hints to the presence of large particles composed of a mixture of icy and dark material.

Overall, the combination of low inversion angle and extreme negative polarisation is unprecedented among comets and asteroids, marking 3I as the first object known with such polarimetric behaviour and possibly representing a previously unobserved class of small planetary bodies. Its polarimetric characteristics provide novel insights into the dust properties of interstellar objects, suggesting that ISOs may encompass a broader diversity than previously recognised. Further post-perihelion observations in the coming months, when the 3I will be visible between phase angles $\sim0$--$30^\circ$ and heliocentric distance $<2$ au, will be critical for fully characterising its polarimetric behaviour and confirming the parallels (or lack there of) with known Solar System bodies, as well as constraining its particle properties based on laboratory and computational studies.

\vspace{5mm}

\begin{small}
\noindent
Based on observations obtained with the FORS2 instrument at the ESO Telescopes at the La Silla Paranal Observatory under program ID 115.29F8.001, 115.29F8.002, 115.29F8.003, 115.29F8.004, 115.29F8.005 (PI Bagnulo). 
Based on observations made with the Nordic Optical Telescope, owned in collaboration by the University of Turku and Aarhus University, and operated jointly by Aarhus University, the University of Turku and the University of Oslo, representing Denmark, Finland and Norway, the University of Iceland and Stockholm University at the Observatorio del Roque de los Muchachos, La Palma, Spain, of the Instituto de Astrofisica de Canarias. The NOT data were obtained under program ID P71-413 (PI Fedorets). The data presented here were obtained in part with ALFOSC, which is provided by the Instituto de Astrofisica de Andalucia (IAA) under a joint agreement with the University of Copenhagen and NOT. Based on observations obtained at BNAO Rozhen during semester 2025A under the program "Investigation of the physico-chemical properties of different classes of comets" led by G. Borisov.
All FORS2 raw data and calibrations are available at {\tt archive.eso.org}.
\end{small}

\section*{Acknowledgements}

Z.G., A.P., and K.M. acknowledge support by the Research Council of Finland (grants \#359893 and \#336546)

R.C.D. and M.G. acknowledge support from grant \#361233 awarded by the Research Council of Finland.

G.B. acknowledges partial support from grant K$\Pi$-06-H88/5 by the Bulgarian National Science Fund.

G.B. gratefully acknowledges observing grant support from the Institute of Astronomy and National Astronomical Observatory, Bulgarian Academy of Sciences.

\bibliography{3Ipol}{}
\bibliographystyle{aasjournal}

\newpage
\appendix 

\section{Instrumental setup and observing strategy}

\begin{figure*}
\begin{center}
\includegraphics[width=0.6\textwidth]{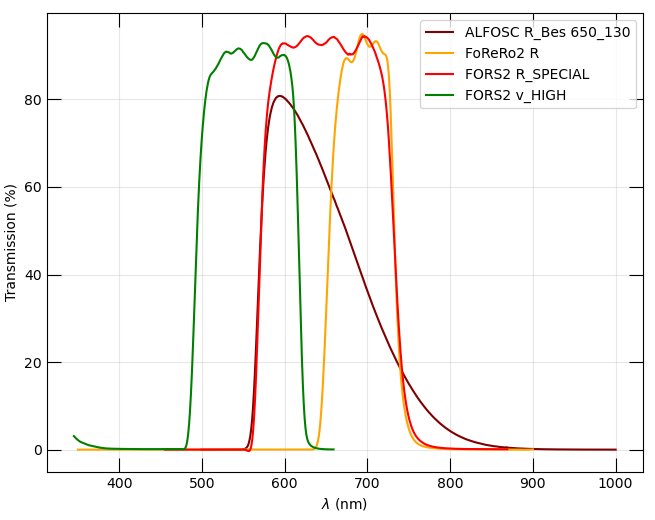}
\end{center}
\caption{\label{Figure3} Transmission curves of R- and V-band filters used for observations. } 
\end{figure*}

Imaging polarimetric observations were obtained in the R- and V-bands using three instruments: FOcal Reducer/low dispersion Spectrograph (FORS2) at the 8.2-m Very Large Telescope (VLT) of the European Sourthern Observatory (ESO; Chile), Alhambra Faint Object Spectrograph and Camera (ALFOSC) at the 2.56-m Nordic Optical Telescope (NOT) in Roque de Los Muchachos (La Palma, Spain), and the two-channel Focal Reducer Rozhen (FoReRo2) at the 2-m Ritchey–Chr\'etien-Coude (RCC) telescope of the Bulgarian National Astronomical Observatory (BNAO) Rozhen. The transmission curves of the filters used are given in Fig.~\ref{Figure3}. 

The polarimetric optics of all three instruments follow the design suggested by \citet{appenzeller1967}: a rotatable retarder waveplate followed by a beam-splitting device. This arrangement produces ordinary and extraordinary beams separated (by $22''$, $15''$ and $62.5''$ for FORS2, ALFOSC and FoReRo2, respectively) on the detector, allowing for simultaneous measurement of orthogonally polarised intensities.

By taking observations consisting of a series of exposures with the $\lambda/2$ retarder waveplate set at position angles $\delta = 0^\circ, 22.5^\circ,45^\circ,...,337.5^\circ$, or a subset of these angles, we measured the degree of linear polarisation using the so-called beam-swapping technique described in \citet{bagnulo2009}. If \fo~and \fe~are the fluxes measured in the parallel and perpendicular beams, respectively, we calculate the reduced Stokes parameters $\pq' = Q/I$ and $\pu' = U/I$ \citep{shurcliff1962} as
% ------------------------------------------- EQUATION 1 -------------------------------------------
\begin{equation}
\px' = {1 \over 2 N } \sum\limits_{j=1}^N \left[ 
\left(\frac{\fo - \fe}{\fo + \fe}\right)_{\delta = \delta_j} - 
\left(\frac{\fo - \fe}{\fo + \fe}\right)_{\delta = \delta_j + 45^\circ}\right]\; ,
\label{Eq_Diff}
\end{equation}
% --------------------------------------------------------------------------------------------------
where $\delta_j \in\ \{0^\circ, 90^\circ, 180^\circ, 270^\circ \}$ for $X=Q$; and $\delta_j \in\ \{22.5^\circ, 112.5^\circ, 202.5^\circ,295.5^\circ \}$ for $X=U$, and $N$ is the number of pairs of exposures per Stokes parameter. We then transform these Stokes parameters to the direction perpendicular to the scattering plane using
% ------------------------------------------- EQUATION 2 -------------------------------------------
\begin{equation}
\begin{array}{rcl}
\pq &=& \phantom{-}\pq'\cos(2\chi) + \pu' \sin (2 \chi) \\
\pu &=&           -\pq'\sin(2\chi) + \pu' \cos (2 \chi) \\
\end{array}
\label{Eq_Rota}
\end{equation}
% --------------------------------------------------------------------------------------------------
where $\chi = \rm{PA} + \Phi + 90^\circ + \epsilon(\lambda)$. Here, PA is the position angle of the instrument (the direction of the principal plane of the beam-splitter) counted counterclockwise from the great circle passing through the comet and the north celestial pole, $\Phi$ is the position angle of the scattering plane (comet-Sun direction relative to comet-north celestial pole direction), and $\epsilon(\lambda)$ is the wavelength-dependent chromatism correction angle of the retarder waveplate.

Regarding observing strategy, we adjusted the PA angle in two distinct ways to accommodate the characteristics of the instruments. In the case of ALFOSC, we rotated the PA to avoid overlap between the extended coma/tail and the photocentre in the \fo~and \fe~beams, estimating the expected direction of the tail in advance using $\Phi$ angle together with the comet's heliocentric velocity vector (listed as PsAng and PsAMV by JPL Horizons). Since both of these values remained close to $90^\circ$ throughout all observations, a simple PA rotation of $90^\circ$ was sufficient. In addition to this, we ensured there was no overlap despite the rotation via visual inspections and numerical tests. 

\begin{table*}
\caption{\label{tableLog2} 3I/ATLAS observing log: $\alpha$ is phase angle, $r$ is heliocentric distance, $\Delta$ is geocentric distance, $\Phi$ is the position angle of the scattering plane, PA is the position angle of the instrument (counted positive N to E), $n_Q$ and $n_U$ are the number of frames used to calculate \pq~and~\pu, respectively. In cases where a column/row parameter is not listed, its value is equal to that in the previous row.} 
%\begin{center}
\centering
\begin{tabular}{cccccccccccc}
\hline\hline
Date & UT & Instr. & $\alpha$ & $r$ & $\Delta$ & $\Phi$ & PA  &Filter& Exp & $n_Q$ & $n_U$ \\ 
     &    &        & (deg)    &(au) &  (au)    & (deg)  &(deg)&      & (s) &     & \\
\hline\hline
2025-Jul-17 & 01:30 & FORS2 & 7.7 & 3.98 & 3.08 & 100.9 &  11.0 & R & 6x70s  & 4 & 2 \\
---         & 01:54 &  ---  & --- & ---  & ---  & ---   & ---   & V & 6x100s & 6 & 0 \\
2025-Jun-18 & 23:57 &  ---  & 8.5 & 3.92 & 3.04 & 100.6 &  11.0 & R & 8x70s  & 6 & 2 \\
2025-Jun-19 & 00:19 &  ---  & --- & ---  & ---  & ---   & 10.5  & V & 6x100s & 6 & 0 \\
---         & 00:51 &  ---  & --- & ---  & ---  & ---   & 10.5  & R & 8x50s  & 8 & 0 \\
2025-Jul-22 & 00:09 &  ---  & 9.7 & 3.82 & 2.98 & 100.2 & 10.2  & R & 6x90s  & 6 & 0 \\
2025-Jul-23 & 20:35 &FoReRo2&10.5 & 3.76 & 2.94 & 100.1 & 0     & R &12x300s & 6 & 6 \\
2025-Jul-24 & 00:37 & FORS2 &10.5 & 3.75 & 2.94 & 100.1 &  10.1 & R & 6x60s  & 6 & 0 \\
---         & 21:00 &FoReRo2&10.8 & 3.73 & 2.93 & 100.1 & 0     & R &24x300s &12 & 12 \\
2025-Jul-29 & 22:27 & ALFOSC&13.0 & 3.56 & 2.84 & ---   & 90.0  & R &16x100s & 8 & 8 \\
2025-Jul-30 & 00:49 & FORS2 & --- & ---  & ---  & ---   & 10.1  & R & 6x80s  & 6 & 0 \\
2025-Aug-12 & 00:22 &  ---  &18.0 & 3.14 & 2.69 & 101.1 & 10.8  & R & 8x40s  & 8 & 0 \\
2025-Aug-15 & 21:37 & ALFOSC&19.3 & 3.01 & 2.66 & 102.0 & 90.0  & R & 16x80s & 8 & 8 \\
2025-Aug-21 & 02:16 & FORS2 &20.8 & 2.85 & 2.62 & 101.9 & 11.8  & R &  8x30s & 4 & 4 \\
---         & 02:27 &  ---  & --- & ---  & ---  & ---   & ---   & V &  4x40s & 4 & 0 \\
2025-Aug-24 & 21:12 & ALFOSC&21.7 & 2.73 & 2.62 & 102.2 & 90.0  & R & 16x80s & 8 & 8 \\
2025-Aug-27 & 18:45 &FoReRo2&22.2 & 2.67 & 2.59 & 102.6 & 0     & R & 8x300s & 4 & 4 \\
2025-Aug-28 & 18:25 &  ---  &22.4 & 2.64 & 2.59 & 102.7 & 0     & R & 8x300s & 4 & 4 \\ 
\hline
\end{tabular}
%\end{center}
\end{table*}

For FORS2, this overlap is not an issue thanks to an additional optical element, the strip-mask (Wollaston mask): composed of strips of opaque material of width equal to the \fo-\fe separation, preventing the superposition of the two images. The resulting image appears as $6.8'\times22''$ strips of double-images, i.e. the ordinary and extraordinary beams neatly separated without overlap, though with the disadvantage of a vignetted FoV. In this case, for most observations, we instead rotated the PA such that the principal axis of the Wollaston prism beam-splitter was aligned perpendicular to the scattering plane, i.e., $\rm{PA} + \Phi \sim 90^\circ$. In this configuration, symmetry dictates $\pq' = \pq$ and $\pu' = \pu = 0$, allowing us to save telescope time by taking exposures with $\delta$ angles corresponding to $X=Q$ only. Since we could not predict the exact timing (and hence, $\Phi$) of the observations, we could not set this alignment perfectly when submitting OBs. We corrected for these small deviations a posteriori by deriving $\pq = \pq' \sec({2\chi})$ (assuming $\pu = 0$), though in practice the correction was negligible (see Table~\ref{tableLog}). In the case of FoReRo2, which also has a Wollaston mask ($55'' \times 435''$), the PA of the instrument was not altered for the observations. 

\section{Data Reduction}
During early epochs, 3I was projected against a crowded Galactic field, requiring careful rejection of frames contaminated by background stars overlapping the comet photocentre and their corresponding pair. After bias subtraction and flat field correction, for each valid exposure, we measured the \fo and \fe~fluxes within circular apertures corresponding to a projected radius of $\sim2000$ km centred on the comet photocentre, subtracting the background sky contribution which we measured in nearby coma/tail-free sky region using sigma-clipping to mitigate the effect of field stars. We opted for this small aperture size to reduce contamination from background stars. We then applied Eqs. \ref{Eq_Diff} and \ref{Eq_Rota} to measure numerical values of linear polarisation. We use this method of aperture polarimetry to compare the bulk properties of 3I to that of other objects. To study the spatial distribution of polarisation within the coma, we construct polarimetric maps from FORS2 data by aligning the whole \fo and \fe~strips according to their photocentres and applying these same equations, and deep imaging maps by simply co-added the aligned strips. We apply a Gaussian filter to the polarimetric maps to reduce noise in the image while preserving the overall structure and edges.

Finally, we observed zero- and high-polarisation standard stars, VICyg12 and BD+28~4211, with ALFOSC to measure and correct for instrumental polarisation, and BD+32 3739 and Hiltner~960 for FoReRo2. This was not necessary for FORS2 as standard star observations are already part of the instrument calibration plan. We calculated the uncertainties of the expressions following standard error propagation methods discussed, for instance, in \citet{bagnulo2009}.

\end{document}